\documentclass{iau}
\usepackage{graphicx}

\title[On the reliability of measuring differential rotation] 
{On the reliability of measuring differential rotation of spotted stars}

\author[ Zsolt K\H{o}v\'ari et al.]   
{Zsolt K\H{o}v\'ari, 
 J\'anos Bartus, 
 Levente Kriskovics, 
 Kriszti\'an Vida \\
 \and Katalin Ol\'ah}

\affiliation{Konkoly Observatory, \\
Konkoly Thege \'ut 15-17., H-1121, Budapest, Hungary \\ email: {\tt kovari}, {\tt bartus}, {\tt kriskovics}, {\tt vida}, {\tt olah@konkoly.hu}}

\pubyear{2013}
\volume{302}  
\pagerange{100--101}
\jname{Magnetic Fields Throughout the Stellar Evolution}
\editors{A.C. Editor, B.D. Editor \& C.E. Editor, eds.}
\begin{document}

\maketitle

\begin{abstract}
Cross-correlation of consecutive Doppler images is one of the most common techniques
used to detect surface differential rotation (hereafter DR) on spotted stars. The disadvantage of a single cross-correlation is,
however, that the expected DR pattern can be overwhelmed by sudden changes
in the apparent spot configuration. Another way to reconstruct the image shear using Doppler imaging is to include
a predefined latitude-dependent rotation law in the inversion code (`sheared image method').
However, special but not unusual spot distributions, such like a large polar cap or an equatorial
belt (e.g., small random spots evenly distributed along the equator),
can distort the rotation profile similarly as the DR does, consequently,
yielding incorrect measure of the DR from the sheared image method. To avoid these problems,
the technique of measuring DR from averaged cross-correlations using time-series Doppler images
('ACCORD') is introduced and the reliability of this tool is demonstrated on artificial data.
\end{abstract}

\firstsection 
\section{Profile distortion by spots vs. differential rotation}

\begin{figure}[b]
\begin{center}
 \includegraphics[width=3.8cm]{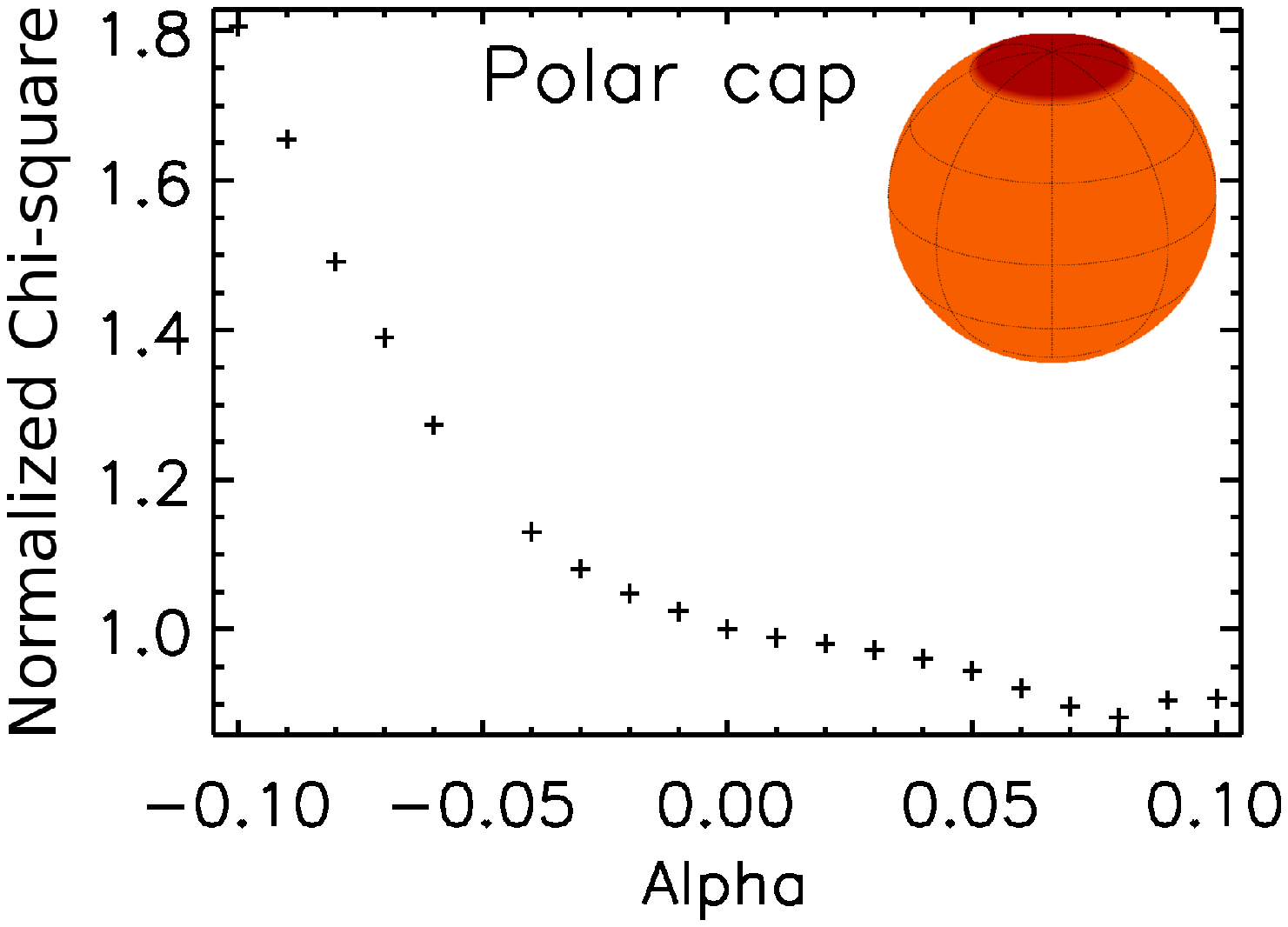}\hspace*{2.0 cm} \includegraphics[width=3.8cm]{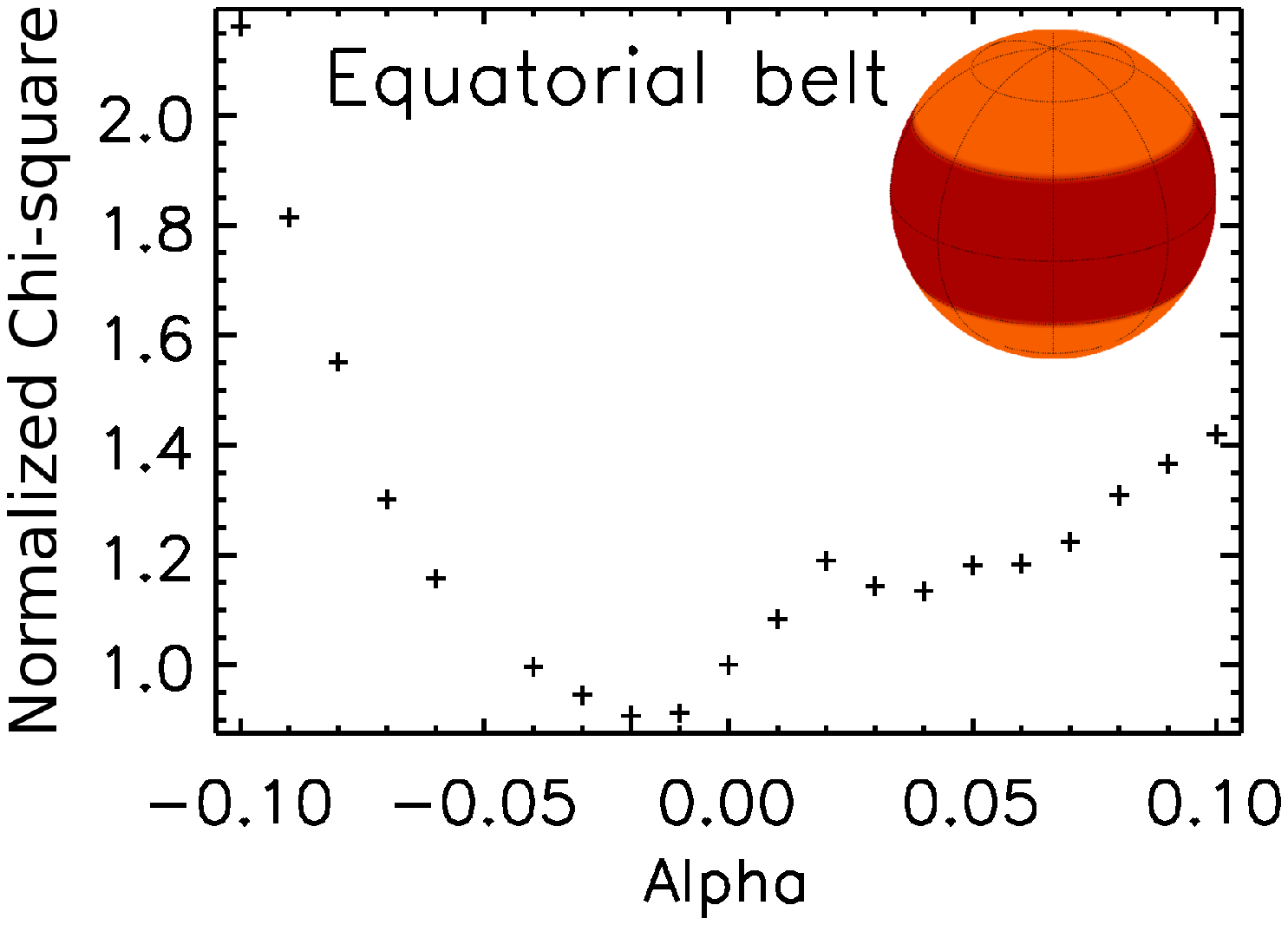} 
 \caption{Biased measure of DR from SIM: a polar cap (left) and an equatorial belt (right) on a rigidly rotating star imitate
solar-type and antisolar-type DR, respectively.}
   \label{fig1}
\end{center}
\end{figure}

Compared to an unspotted star, the theoretical shape of a broadened spectral line profile
from a spotted star is perturbed
by starspots and surface differential rotation (DR). Indeed, in some cases
these two effects cause very similar distortions, thus, their separation
can be difficult, if possible at all. This difficulty may result in biased observation of surface DR, when applying
the `sheared image method' (hereafter SIM, \cite[Weber 2004]{Weber04}), where the line profile fit is minimized
for a given range of the image shear parameter $\alpha$. In Fig.\,\ref{fig1} we demonstrate
how a polar cap/equatorial belt on a rigidly rotating star imitates
solar-type/antisolar DR, respectively. That is, assuming realistic conditions,
SIM yields better
fits for non-rigid solutions for these two special examples (i.e., $\alpha$ of $0.08$ and $-0.02$ instead of zero for the supposed rigid rotation).
We note, that the line profile distortions due to polar cap/equatorial belt
are similar in shape to those that are attributed to the
respective solar/antisolar DR (cf. \cite[Reiners \& Schmitt 2002]{ReSc02}). In Sect.\,2 we give a tool to resolve this problem.

\begin{figure}[tb]
\begin{center}
 \includegraphics[width=3.5cm]{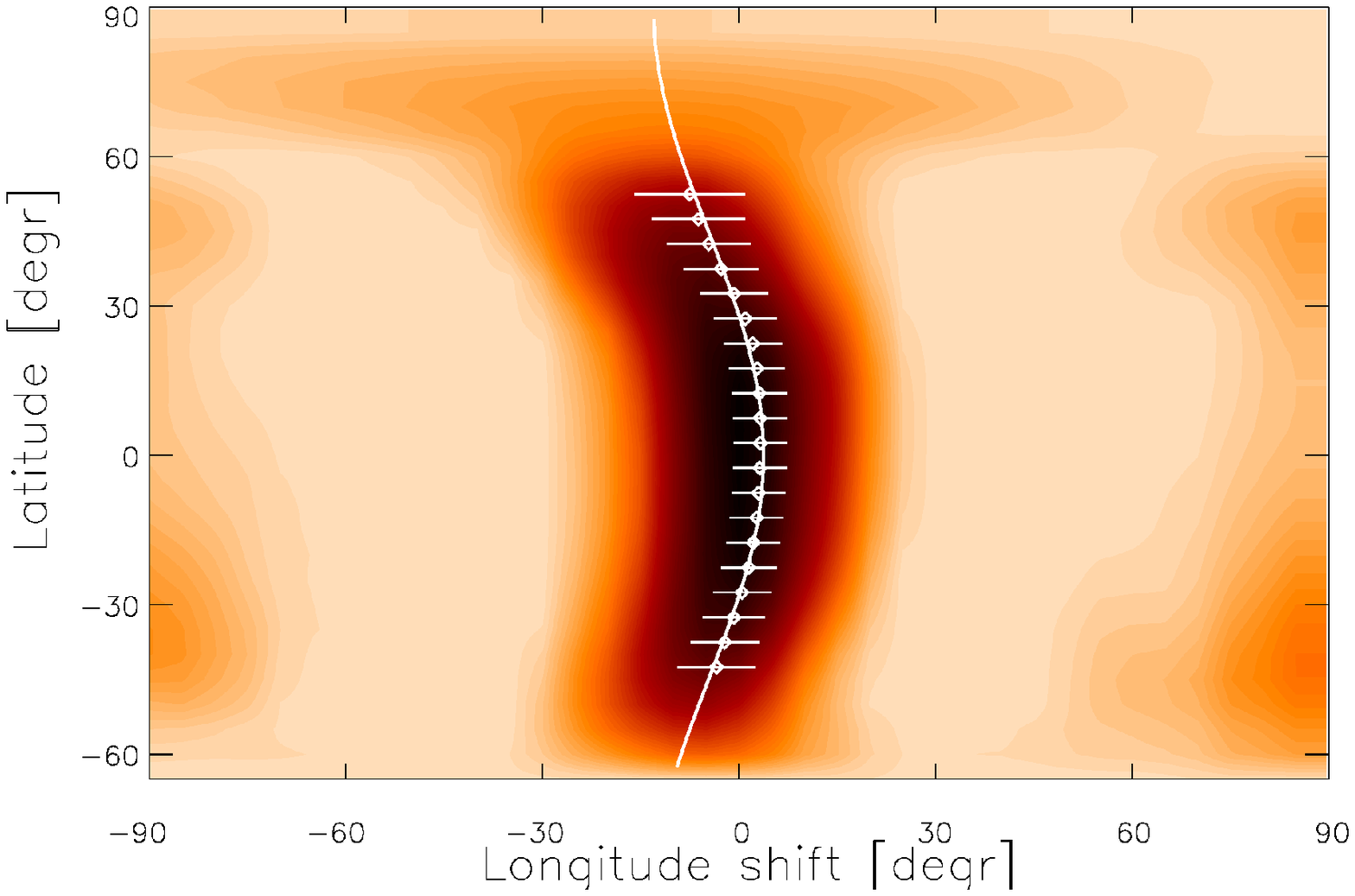}\hspace*{1.0 cm} \includegraphics[width=3.5cm]{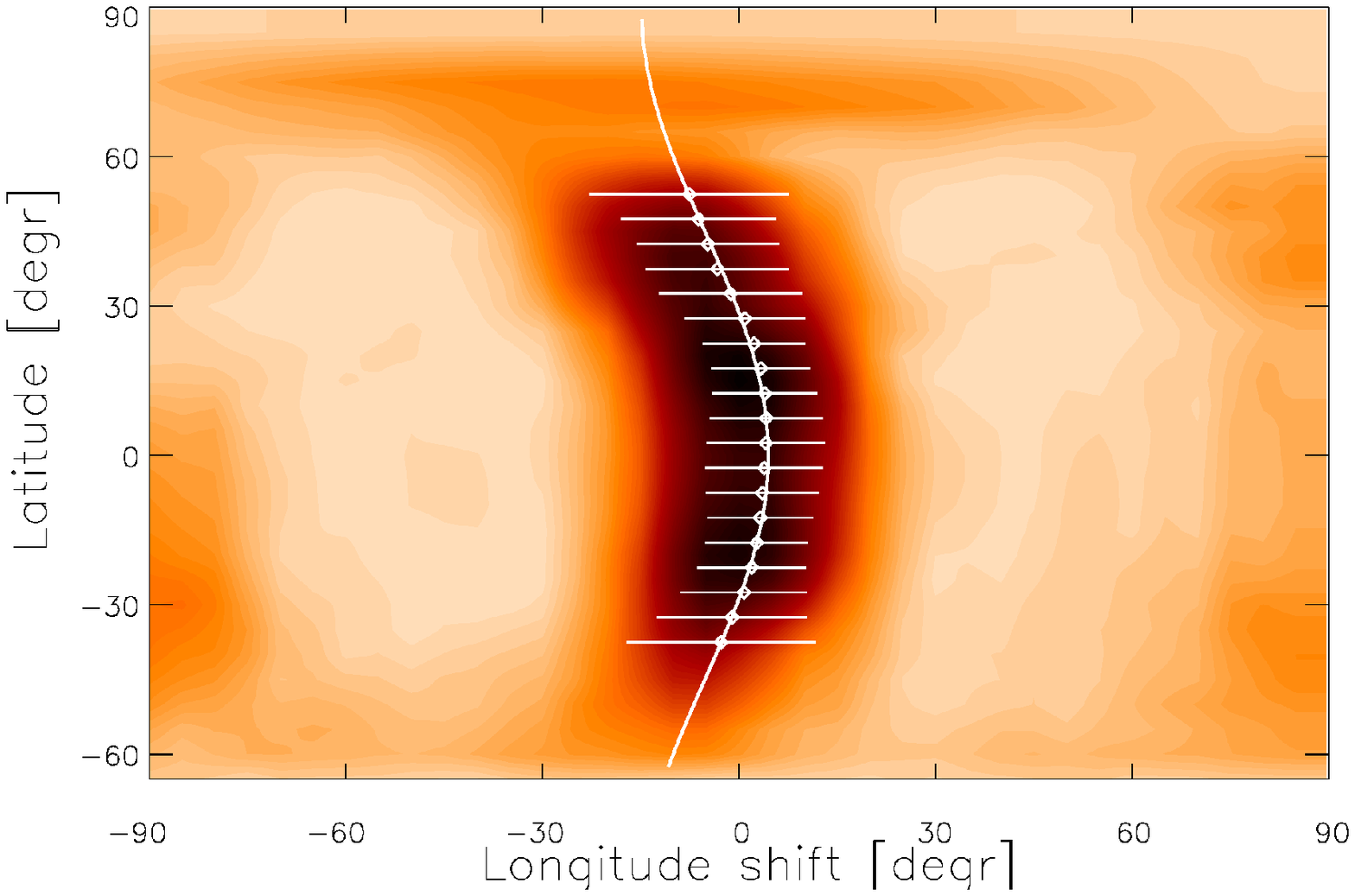} \hspace*{1.0 cm} \includegraphics[width=3.5cm]{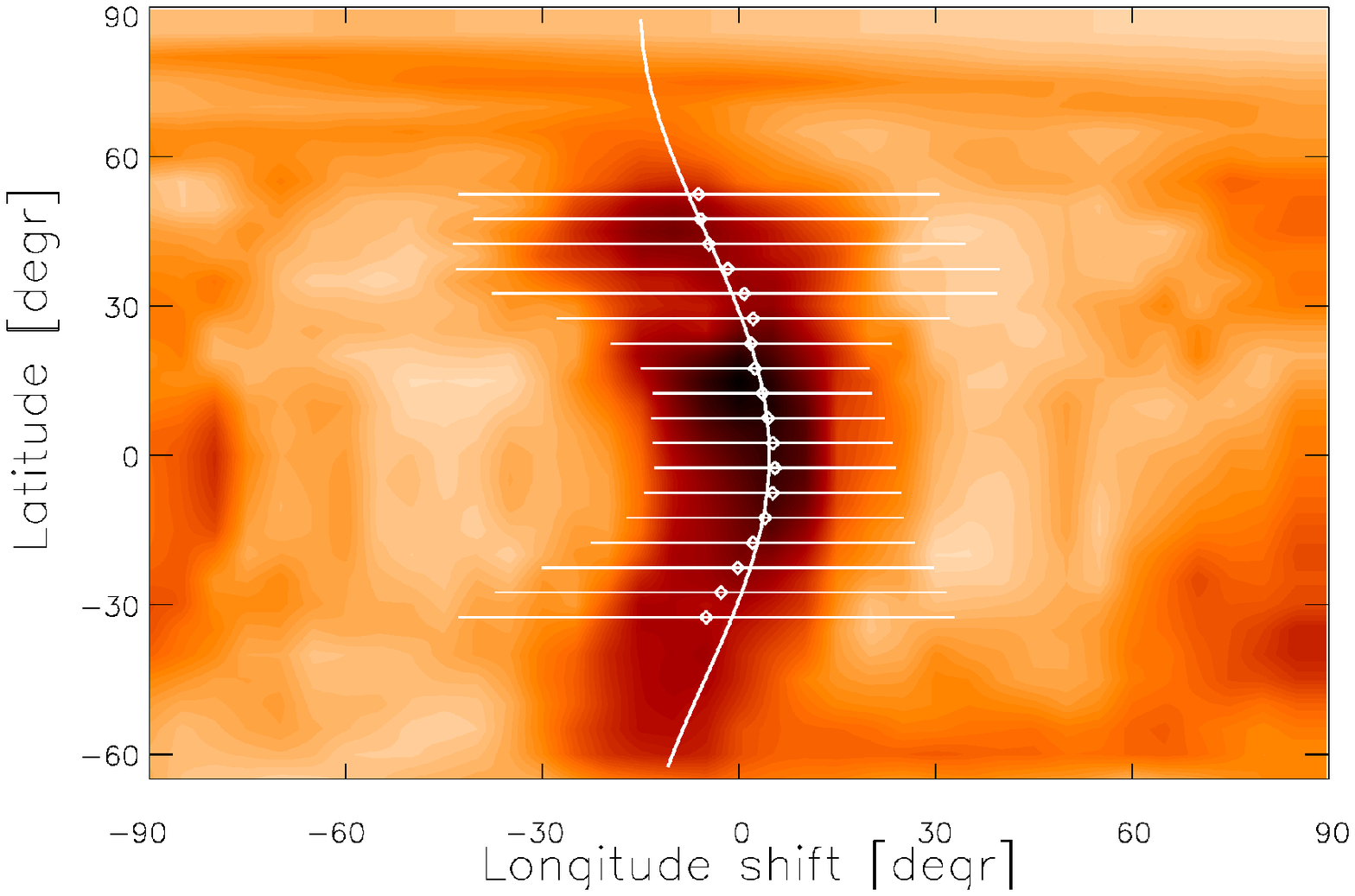} 
 \caption{Averaged cross-correlations reveal DR laws with correct $\alpha$ parameters close to the original one of $0.006$, either when assuming no noise (left), or assuming S/N=200 (middle) and S/N=100 (right), yielding $\alpha=0.0058\pm0.0004$, 
$\alpha=0.0067\pm0.0033$ and $\alpha=0.0068\pm0.0048$, respectively. Note that uncertainty comes also from imperfect image reconstruction due to the necessarily incomplete phase coverage,
thus errors are expected even when no noise is added.}
   \label{fig2}
\end{center}
\end{figure}

\section{Measuring surface DR from average cross-correlations}

Cross-correlation
of two consecutive 'snapshots' of the stellar surface (i.e., Doppler images) can reveal
the DR pattern (\cite[Donati \& Collier Cameron 1997]{DoCo97}). The advantage of this method, as against the sheared image, is that no
predefined rotation law is assumed. The disadvantage of a {\it single} cross-correlation is,
however, that the DR pattern in the correlation map is often overwhelmed by sudden changes
in the apparent spot configuration.
Our method called ACCORD (acronym from 'Average Cross-CORrelation of consecutive Doppler
images) is based on averaging as many correlation maps as possible
to suppress the effect of such stochastic spot changes (see \cite[K\H{o}v\'ari et al. 2004, ]{kovetal04}\cite[2012]{kovetal12} for
details). To demonstrate the reliability of this method we generated artificial time-series
observations from a test star with the stellar parameters of the fast rotating active dwarf LQ\,Hya.
Even the data sampling (phase coverage) followed our actual 70-night long observing run in
1996/97 (\cite[K\H{o}v\'ari et al. 2004]{kovetal04}). We place a polar cap, as well as high- and low latitude spots
on the differentially rotating surface with an assumed surface shear $\alpha$ of 0.006 (i.e., a weak
solar-type DR, cf. \cite[K\H{o}v\'ari et al. 2004]{kovetal04}). Visit {\tt www.konkoly.hu/solstart/misc/testDR.gif}
to see the surface evolution of the test star. After generating artificial spectra (for a given
noise level) we followed the data handling as described in \cite[K\H{o}v\'ari et al. (2004)]{kovetal04}.
Fig.\,\ref{fig2} demonstrates how ACCORD reveals the correct DR law.
Contrarily, {\it single} cross-correlations
resulted in much scattered $\alpha$ values with relative errors of 23\% (no noise), 48\% (S/N=200) and 150\% (S/N=100).

Finally, we performed a test assuming rigid rotation with either a polar cap or an equatorial belt, as described in Sect.~1, to see how specious are these configurations for ACCORD. Again,
this method proved to be sufficiently robust, resulting in basically the expected zero shear for
both cases: $\alpha=0.000\pm0.001$ for the polar cap and $\alpha=0.003\pm0.002$ for the equatorial belt, i.e., far better results than the respective values of $0.08$ and $-0.02$ from SIM (cf. Fig.\,\ref{fig1}). 


\begin{acknowledgments}
This work has been supported by the Hungarian Science Research Program OTKA K-81421,
the “Lend\"ulet-2009” and “Lend\"ulet-2012” Young Researchers’ Programs of the Hungarian Academy
of Sciences and by the HUMAN MB08C 81013 grant of the MAG Zrt.\end{acknowledgments}

\end{document}